%% file: research_paper.tex
\documentclass[a4paper,UKenglish,cleveref,autoref,thm-restate]{lipics-v2021} %anonymous
\pdfoutput=1

\usepackage{cite}
\usepackage{amsfonts}
\usepackage{algorithmic}
\usepackage{makecell}
\usepackage{url}
\usepackage{xurl}
\usepackage{tikz}
\usetikzlibrary{arrows.meta, positioning}
\usepackage[most]{tcolorbox}

\title{Investigating Notable Metadata Practices in PyPI Libraries: An Empirical Study about Repository and Donation Platform URLs}

\titlerunning{Investigating Notable Metadata Practices in PyPI Libraries}

\author{Alexandros Tsakpinis}
{fortiss GmbH, Munich, Germany}
{tsakpinis@fortiss.org}
{https://orcid.org/0000-0001-6561-2866}
{}

\author{Nicolas Raube}
{Technical University of Munich, Munich, Germany}
{nicolas.raube@tum.de}
{https://orcid.org/0009-0006-1803-7604}
{}

\author{Alexander Pretschner}
{Technical University of Munich, Munich, Germany}
{alexander.pretschner@tum.de}
{https://orcid.org/0000-0002-5573-1201}
{}

\authorrunning{A. Tsakpinis, N. Raube, and A. Pretschner}

\Copyright{Alexandros Tsakpinis, Nicolas Raube, and Alexander Pretschner}



\relatedversion{}

\acknowledgements{}

\hideLIPIcs % For Preprints on Arxiv

\newcommand{\numRaters}{23 }

\newcommand{\tableWidth}{1}

\begin{document}
\nolinenumbers
\maketitle

\input{sections/0_Abstract}
\input{sections/1_Introduction}
\input{sections/2_Background_and_Related_Work}

\input{sections/3_Methodology}
\input{sections/4_Results}
\input{sections/5_Discussion}
\input{sections/6_Threats_to_Validity}
\input{sections/7_Conclusion_and_Future_Work}
\input{sections/8_Data_Availability}

\bibliography{research_paper}

\end{document}

%% file: sections/0_Abstract.tex
\begin{abstract}
    \textbf{Background:} 
    Open source software (OSS) libraries are critical components of modern software systems, yet their metadata---particularly links to source code repositories and donation platforms---is often incomplete, outdated, or inconsistent. Such deficiencies hinder dependency monitoring, security assessment, and the sustainability of OSS projects.
    \textbf{Aims:} 
    This study aims to explain notable metadata practices in PyPI libraries, focusing on platform dominance, outdated links, and missing references to repositories and donation platforms. As this investigation relies on large-scale qualitative survey data, we further evaluate the robustness and quality of the LLM-based topic modeling approach used to derive the findings.
    \textbf{Method:} 
    We conducted two surveys targeting PyPI authors and maintainers, collecting 1,776 open-ended responses. To analyze these responses, we developed a LLM-based topic modeling pipeline using LLaMA 3.3 70B, including preprocessing, topic extraction, and topic merging. Robustness was assessed across 30 repeated runs using Jaccard and cosine similarity, while topic quality was evaluated by 23 experts using a structured assessment framework and Randolph’s Kappa.
    \textbf{Results:} 
    The findings reveal that missing or outdated repository links are primarily associated with oversight, lack of awareness, or perceived irrelevance, while platform dominance is driven by ideological, technical, and organizational factors. Donation platform links are often omitted due to skepticism, limited perceived benefit, or lack of knowledge, and are preferentially placed on GitHub for visibility reasons. The topic modeling approach demonstrated high robustness (up to 88\% lexical and 92\% semantic similarity) and produced high-quality topics, with approximately 77--78\% meeting all evaluation criteria and moderate inter-rater agreement~($\kappa\approx0.55$).
    \textbf{Conclusions:}
    Metadata-related issues in PyPI stem from a combination of usability limitations, insufficient guidance, and maintainer behavior. Improving platform support through clearer guidance, better interfaces, and automated validation can enhance metadata completeness, benefiting security and OSS sustainability. Furthermore, our results suggest that LLM-based topic modeling can serve as a reliable and scalable method for analyzing short-text survey data.
\end{abstract}

%% file: sections/1_Introduction.tex
\section{Introduction}

%\textbf{Situation:} 
Open Source Software (OSS) has become a fundamental part of the product life cycle, driven by commercial, engineering, and quality considerations~\cite{ebert2008open}. Today, OSS components constitute 80--90\% of the code in commercial products, highlighting their widespread adoption~\cite{pittenger2016open, oss2022}. A crucial subset is software libraries, often called packages, which enable reuse of well-tested functionality and reduce development effort~\cite{bauer2012structured, cox2025fifty}. Once incorporated, a library becomes a dependency~\cite{cox2019surviving}. However, dependencies introduce risks, including security vulnerabilities (e.g., Log4j~\cite{log4j2021}) and supply chain attacks (e.g., LiteLLM~\cite{litellm}, XZ Utils~\cite{xz2024}), posing a significant challenge to the software industry~\cite{decan2018impact}. While communities typically release fixing patches promptly~\cite{rahkema2022swiftdependencychecker}, discontinued or inactive projects pose significant challenges due to missing maintenance~\cite{bauer2012structured, raemaekers2011exploring, tsakpinis2023analyzing}. Monitoring maintenance activity requires valid repository links, for example when using tools like the OpenSSF Scorecard~\cite{openssf_scorecard}. Taking the PyPI ecosystem as an example---which stands out due to its widespread use in modern software development and the central role libraries play in Python’s development practices~\cite{decan2016topology, abdalkareem2020impact, Octoverse2025}---maintainers can link repositories (e.g., GitHub or GitLab) in the “Project Links” section~\cite{devtools_pypi}. In parallel, funding is a key enabler of OSS maintenance~\cite{tidelift2024, medappa2023sponsorship}. Donation platforms such as GitHub Sponsors or Open Collective provide channels for financial support, directly influencing project sustainability~\cite{zhang2025exploring}. Making donation links visible on platforms like PyPI can lower barriers for potential sponsors and increase support, whereas missing links reduce visibility and thus funding opportunities. Moreover, corporate programs by Microsoft, Stripe, and Indeed often require registration on such platforms~\cite{microsoft_foss, stripe_oss, indeed_foss}, effectively making them a prerequisite for participation. To increase visibility among potential supporters, maintainers can link donation platforms via PyPI’s “Project Links” or GitHub’s \texttt{FUNDING.yml}~\cite{funding_yml}.

%\textbf{Problem:}
However, previous research highlights that library authors and maintainers frequently neglect to include links to repositories in the "Project Links" section on PyPI, while donation platform links are often omitted from  PyPI project pages and associated GitHub repositories~\cite{tsakpinis_pretschner_2024, tsakpinis2025analyzing}. Our replication of these studies using data from December 2025 confirms this trend~\cite{replication_package}: only 65.8\% of all PyPI libraries provide a link to their code repository, while 5.9\% reference a donation platform. The analysis also reveals several broader issues within the PyPI ecosystem. First, outdated URLs are common, affecting 12.9\% of all GitHub repository links and 35.8\% of donation platform links to GitHub Sponsors. Second, GitHub dominates as the preferred source code management (SCM) system, accounting for 97.2\% of repository links, while alternatives like GitLab, Bitbucket, and SourceHut collectively make up just 2.8\%. Similarly, most donation platform URLs (96.7\%) are not listed in PyPI's "Project Links" section but instead appear exclusively on GitHub repositories. Despite the clear benefits of linking repositories and donation platforms, their frequent omission and reported issues raise important questions. We seek to understand why authors and maintainers choose to omit such links. Additionally, we aim to explore why URLs are not kept up to date and why GitHub dominates over alternative SCMs. Finally, we investigate why maintainers prefer to list donation platforms on GitHub repositories rather than on their PyPI project pages. To guide our empirical analysis, we formulate the following two research questions (RQs):

\textbf{RQ1:} What factors are associated with notable metadata practices concerning repository URLs in PyPI libraries, specifically platform dominance, outdated and missing links?

\textbf{RQ2:} What factors are associated with notable metadata practices concerning donation platform URLs in PyPI libraries, specifically platform dominance, outdated and missing~links?

%\textbf{Solution:}

In order to identify the factors associated with notable metadata practices related to repository and donation platform URLs in PyPI libraries, we conducted two independent surveys with randomly selected PyPI authors and maintainers. Based on the replicated empirical findings~\cite{replication_package}, we designed open-ended survey questions to encourage creativity and reveal topics we would not have thought about when providing closed-ended questions~\cite{hahn2024improving}. Given the large volume of responses---totaling 1,776 answers---we decided on an automated approach to extract relevant insights. To this end, we developed a topic modeling pipeline based on a Large Language Model (LLM) to automatically identify key topics in the responses. To validate the robustness of our approach, we repeated the experiment 30 times and measured the lexical and semantic consistency of the resulting topics across all runs using Jaccard and Cosine similarity metrics. Finally, to assess the quality of the extracted topics, we generated an evaluation form based on the topic modeling results, which was then distributed across \numRaters industry and research experts familiar with the broader topic around OSS libraries.

%\textbf{Contribution:} 
This paper makes two main contributions. First, we provide empirical insights into the metadata practices of PyPI libraries by identifying factors associated with missing, outdated, and platform-dominated repository and donation platform links. Our findings uncover practical, organizational, and ideological barriers that influence maintainers’ decisions, offering actionable implications for improving metadata quality and support within package registries such as PyPI. Second, we evaluate our LLM-based topic modeling approach for analyzing large-scale survey data. While the underlying techniques build on existing work, we provide a systematic assessment of their robustness and topic quality in this context. Our results show high consistency across repeated runs (88\% lexical and 92\% semantic similarity) and moderate agreement among 23 expert raters (Randolph’s $\kappa = 0.55$), providing empirical indications of a reliable and scalable alternative to manual qualitative analysis.

%% file: sections/2_Background_and_Related_Work.tex
\section{Background and Related Work}

\subsection{Library Metadata}

OSS package registries expose metadata to support activities such as discoverability and dependency management. Prior work has leveraged this metadata to characterize registries at scale, reporting trends in releases, dependencies, licenses, and contributors~\cite{bommarito2019empirical}. However, metadata quality remains challenging even for widely used fields such as licensing, with irregularities and incomplete information observed across package registries including~PyPI~\cite{wu2024large}. 

A similar challenge arises for repository metadata. As repository links are not mandatory and may appear in different metadata locations, Vu proposed py2src, a tool that combines multiple information sources including PyPI metadata to identify GitHub repository URLs and provide reliability indicators for the returned links~\cite{vu2021py2src}. Repository mapping is also essential for studying discrepancies between source repositories and published artifacts: LastPyMile demonstrates that differences between PyPI artifacts and corresponding source repositories are pervasive and can introduce operational and security risks, underscoring the importance of correct package--repository links~\cite{vu2021lastpymile}. Similar concerns have been studied in other ecosystems: for instance, Goswami et al.\ investigate the reproducibility of npm packages and identify metadata- and toolchain-related factors behind non-reproducible build artifacts~\cite{goswami2020investigating}. Tsakpinis and Pretschner explored how libraries and dependencies in the PyPI ecosystem link to GitHub repositories by analyzing URLs included in the "Project Links" section. They identified notable empirical findings that motivated further investigations in the present study~\cite{tsakpinis_pretschner_2024}. Addressing the issue of missing repository URLs for PyPI libraries, Gao et al.\ proposed PyRadar, a framework for automatically retrieving and validating repository information. Their work also suggested improvements, such as linking package registries with SCM systems to enhance repository validation and introducing warnings for incorrect package metadata~\cite{gao2024pyradar}, in line with the findings of Tsakpinis and Pretschner~\cite{tsakpinis_pretschner_2024}.

Beyond repository links, prior research has examined the use of donation platform URLs in individual PyPI libraries and their dependencies, a study that was later replicated as a motivation for this research. This analysis examined the use of common donation platforms (e.g., GitHub Sponsors, Open Collective, Patreon) based on data from the "Project Links" section on PyPI and GitHub's \texttt{FUNDING.yml} files, which provide a standardized mechanism for advertising funding options on GitHub~\cite{tsakpinis2025analyzing}. Complementing these registry-centric analyses, GitHub Sponsors has been studied directly to characterize participation patterns, maintainer motivations, and perceived shortcomings of this platform-specific funding mechanism~\cite{shimada2022githubsponsors, zhang2022sponsor}. Overney et al.\ investigated how donation platforms appear in GitHub repositories of NPM libraries by mining README files, though their work was conducted before GitHub Sponsors was introduced~\cite{overney2020not}. A survey by Tidelift reports that roughly 25\% of independent maintainers receive support through donation platforms, but provides little detail on why many maintainers do not advertise funding links~\cite{tidelift2024}. To gain deeper insight into the effectiveness of such platforms, Zhang et al.\ studied Open Collective and found significant short-term, but no significant long-term effects on development and maintenance activity~\cite{zhang2025exploring}. Finally, visibility effects have been observed for GitHub Sponsors: mentions of Sponsors profiles on Twitter/X are associated with changes in sponsorship activity, supporting the idea that maintainers may place funding links where they expect higher discoverability~\cite{fan2024mygithubsponsors}.

More broadly, these metadata-related challenges highlight that OSS ecosystems are not purely technical infrastructures, but socio-technical systems shaped by interactions between technical artifacts, developer practices, and community and governance processes~\cite{feng2026charting}.
While prior work focuses on detecting, quantifying, or resolving metadata issues~\cite{gao2024pyradar, tsakpinis_pretschner_2024, tsakpinis2025analyzing}, no study has examined the factors associated with notable metadata practices in PyPI libraries related to code repositories and donation platforms, leaving a gap we aim to address.

\subsection{LLM-based Topic Modeling}

Topic modeling is a natural language processing technique, typically based on statistical modeling or machine learning, that uncovers latent topical patterns in unstructured text collections~\cite{egger2022topic, sharifian2022analysing}. A topic is a set of semantically related words, and the goal is to extract topics that accurately and coherently represent a document collection~\cite{churchill2022evolution}. It is widely used for tasks such as document summarization, text segmentation, and text matching, particularly in contexts like social media or product reviews~\cite{abdelrazek2023topic}. Automated approaches are preferred over manual coding as they reduce errors, bias, and inconsistencies~\cite{pietsch2018topic, roberts2014structural}, especially for large datasets~\cite{chung2022text}. Traditional probabilistic models have been complemented by embedding-based methods operating in semantic vector spaces, with a shift toward neural topic models using word embeddings~\cite{abdelrazek2023topic}. This evolution led to approaches such as BERTopic, combining contextual embeddings with clustering to produce coherent and interpretable topics~\cite{grootendorst2022bertopic}. Such methods have been successfully applied to software engineering datasets, including Stack Overflow posts and GitHub discussions~\cite{akbarpour2025unveiling, vaccargiu2024sustainability}. LLMs further extend these approaches by assisting in interpreting and labeling topics, often producing labels aligned with human judgment~\cite{rijcken2023towards}, and preferred over those generated only by traditional methods~\cite{li2023can}.

Unlike traditional topic modeling methods that estimate distributions or cluster documents, recent work increasingly adopts LLM-based approaches that generate topics directly. These approaches leverage the model’s training data, the input corpus, and a given prompt. Documents can be processed individually or in batches, with batching being more efficient as the prompt is only executed once per batch. LLM-based topic modeling typically relies on a system prompt defining the task and output format, and an user prompt providing the documents. Since LLMs may produce semantically similar topics with different names, a merging step is often applied via prompt-based matching to consolidate related topics~\cite{mu2024large}. While effective for short-text datasets and often aligned with human evaluations, the prompt-based topic merging step may merge unrelated topics or alter granularity due to limited~context~\cite{wang2023prompting}.

LLM-based methods have been shown to generate topics with high consistency and distinctiveness~\cite{doi2024topic}, and can outperform traditional approaches such as BERTopic on short texts according to human evaluation~\cite{wang2023prompting}. Prior work identifies GPT-4-based methods as state-of-the-art~\cite{doi2024topic}, although results may evolve with newer models. While GPT models tend to follow instructions more strictly than open models~\cite{mu2024addressing}, they offer limited transparency and control~\cite{khandelwal2024investigating}. In contrast, Llama (13B) achieves comparable quality, coherence, and even greater topic diversity, while enabling local execution for processing confidential data~\cite{wang2023prompting}. The model’s context window must be sufficiently large to avoid information loss~\cite{khandelwal2024investigating}, although this is less critical for short-text data. Despite their advantages, LLM-based approaches introduce challenges. One key issue is hallucination, where generated topic words do not appear in the source corpus, potentially reducing factual accuracy. However, such cases are typically rare and often involve semantically related terms that do not hinder interpretation~\cite{doi2024topic}. Another challenge is controlling topic granularity, as LLMs may not consistently follow abstraction constraints. Providing task-specific context or seed topics can help mitigate this issue~\cite{wang2023prompting}.

Assessing the robustness of LLM-based topic modeling is essential to ensure that identified topics represent stable themes rather than random artifacts~\cite{wagner2024towards}. Prior work recommends repeated model executions and comparison of resulting topics using similarity measures such as Jaccard or cosine similarity~\cite{siska2024examining}. 
While classical models represent topics as word lists that can be assessed using coherence metrics, LLMs typically generate topic names directly, making such metrics inapplicable. Consequently, human evaluation is commonly required to assess the interpretability and relevance of generated topics~\cite{mu2024addressing, mu2024large}.

%% file: sections/3_Methodology.tex
\section{Methodology}
This study follows an empirical research design combining qualitative data collection with automated analysis and systematic evaluation, as described in the following sections.

\subsection{Survey Design}
We developed our survey questions based on replicated empirical findings from prior research~\cite{replication_package, tsakpinis_pretschner_2024, tsakpinis2025analyzing}. We conducted two separate surveys---one focusing on repository URLs and one on donation platform URLs---to keep each questionnaire concise and focused, and to reduce cognitive load for participants. We formulated questions to explore the factors associated with notable patterns observed in the quantitative data. The responses to these questions reveal topics that directly contribute to answering specific research questions. This ensures a structured mapping: from a replicated finding in related work to a corresponding survey question~(\textbf{SQ}), which links to one research questions~(\textbf{RQ}).
To conduct our surveys, we used Microsoft Forms and ensured all responses were anonymized. We provided participants with an optional field to enter their email address if they wished to be notified about the survey results. Additionally, we included a binary question asking whether the provided email could be used for future contact regarding related research. Both surveys are included as PDFs in our replication package, while this paper presents only a condensed version of their content~\cite{replication_package}.
In the following, we outline the mapping between empirical findings, survey questions, and research questions, highlighting the research gaps we aim to address.

\subsubsection{RQ1: Survey on Repository URLs}
Findings show that GitHub is the most dominant SCM system, with 97.2\% of repository URLs pointing to GitHub, while the remaining 2.8\% are distributed among alternatives like GitLab or BitBucket~\cite{tsakpinis_pretschner_2024}. To understand why maintainers choose alternative platforms over the widely-used GitHub option, we include \textbf{SQ-1.1:}~\textit{If you assigned a GitLab, Bitbucket, Gitea, Codeberg or sourcehut URL to your PyPI library, what were the reasons for not hosting the code on GitHub and linking it accordingly?}
Additionally, results reveal that 12.9\% of assigned GitHub repository URLs are outdated, preventing dependent projects from monitoring their activities~\cite{tsakpinis_pretschner_2024}. To investigate why maintainers do not update these URLs, we ask \textbf{SQ-1.2:}~\textit{Can you please (double) check that your assigned URL is still working. If it returns an error code 404, why haven't you updated the URL accordingly?}
Finally, findings indicate that 22.5\% of PyPI libraries do not provide any repository URL~\cite{tsakpinis_pretschner_2024}, limiting the ability of dependent projects to monitor activities. Therefore, we include \textbf{SQ-1.3:}~\textit{If you have not assigned any URL to your PyPI library, what were the reasons for this?}

\subsubsection{RQ2: Survey on Donation Platform URLs}
The quantitative analysis revealed that only 3.3\% of donation platform links are present on PyPI project pages, while 96.7\% are found exclusively on GitHub repositories~\cite{tsakpinis2025analyzing}. To better understand why maintainers omit donation platform URLs on PyPI, we introduced \textbf{SQ-2.1:}~\textit{If you assigned a link to any donation platform exclusively on the GitHub repository, but not on the library's PyPI project page, what were the reasons for this?}
Additionally, the results showed that 35.8\% of all identified GitHub Sponsors links were outdated~\cite{tsakpinis2025analyzing}, preventing potential supporters from contributing financially. To explore this issue, we include \textbf{SQ-2.2:}~\textit{If you have linked a GitHub Sponsors profile on your library's PyPI project page, but the URL is not reachable (status code 404) anymore, what are the reasons for this?}
Finally, findings show that 94.0\% of libraries do not include links to any donation platform~\cite{tsakpinis2025analyzing}, limiting opportunities to receive financial support. To investigate this, we included \textbf{SQ-2.3:}~\textit{If you have not assigned a link to any donation platform on the library's PyPI project page or on the GitHub repository, what were the reasons for this?}

\subsection{Identification of Survey Participants}
Survey participants were identified by retrieving a list of all available PyPI libraries from an endpoint~\cite{pypi-simple} and, where available, extracting author or maintainer email addresses from an additional endpoint~\cite{pypi-json}. The contact details are voluntarily provided by their authors and maintainers when packaging and publishing a package on PyPI~\cite{pypi_metadata_tutorial}. To ensure compliance with PyPI’s policies regarding the scientific use of this contact information, we reached out to a representative of the Python Packaging Authority (PyPA)---the working group responsible for maintaining core packaging infrastructure and tools within the Python ecosystem, of which PyPI is a central component~\cite{pypa_website}. When distributing the survey via email, we ensured a concise and respectful tone to minimize disruption for recipients who chose not to participate. To preserve anonymity and minimize barriers to participation, we deliberately did not collect detailed demographic or expertise-related information about respondents.

\subsection{Data Analysis}
We implemented a modular data science pipeline with sequential steps for preprocessing the raw survey data, performing topic modeling, calculating robustness measures across repeated runs, and generating an evaluation form including logic to assess the evaluation responses. Our resulting pipeline is illustrated in Figure~\ref{fig:topic_modeling_pipeline}.

\begin{figure}[ht]
    \centering
    \resizebox{\columnwidth}{!}{%
        \begin{tikzpicture}[
            node distance=1.8cm, 
            block/.style={
                draw, 
                rectangle, 
                minimum height=1.2cm, 
                minimum width=2.6cm, 
                align=center,
                font=\small\bfseries,
                line width=0.8pt,
                rounded corners=2pt,
                fill=gray!5
            },
            arrow/.style={
                -{Stealth[scale=1.3]}, 
                line width=0.8pt
            },
            label/.style={
                font=\scriptsize,
                above,
                inner sep=4pt,
                align=center
            }
        ]

        % Nodes
        \node (clean) [block] {Clean\\Responses};
        \node (llm)   [block, right=of clean] {LLM-based\\Topic Modeling};
        \node (prep)  [block, right=of llm] {Prepare\\Evaluation Data};
        \node (form)  [block, right=of prep] {Create\\Evaluation Form};

        % Input Arrow
        \draw [arrow] ([xshift=-1.8cm]clean.west) -- (clean.west) 
            node [midway, label] {Raw\\Responses};

        % Connecting Arrows
        \draw [arrow] (clean) -- (llm) 
            node [midway, label] {Cleaned\\Responses};

        \draw [arrow] (llm) -- (prep) 
            node [midway, label] {Identified\\Topics};

        \draw [arrow] (prep) -- (form) 
            node [midway, label] {Evaluation\\Data};

        % Output Arrow
        \draw [arrow] (form) -- ([xshift=1.8cm]form.east) 
            node [midway, label] {Evaluation\\Form};

        \end{tikzpicture}
    }
    \caption{Overview of the data analysis pipeline, showing the transformation of raw responses into extracted topics and evaluation artifacts. The robustness assessment is omitted for visual clarity.}
    \label{fig:topic_modeling_pipeline}
\end{figure}

\subsubsection{Data Preprocessing}

The first step of the topic modeling pipeline cleans the raw survey answers by removing noise. Noise in this dataset primarily consists of a variation of "not applicable" answers (e.g., "N/a", "* N.A.", "Not applicable.", "pass") or single characters expressing the same meaning (e.g., ".", "-"), which do not provide any relevant information for answering the survey questions. Exactly matching occurrences of the irrelevant expressions are replaced with a standardized "not applicable" placeholder so that subsequent processing steps can easily filter them out.

\subsubsection{LLM-based Topic Modeling}
The LLM-based topic modeling approach is applied on the cleaned survey responses, which have an average length between 80 and 95 characters across both surveys. This places them well within the original tweet limit of 140 characters~\cite{x_counting_characters}, allowing them to be treated as short documents. Prior work shows that LLM-based topic modeling performs particularly well on such short-text data~\cite{wang2023prompting}, making our dataset well-suited for this approach. 

We process documents in batches of 10 and merge redundant topics through prompt-based matching. While the context window would allow for larger batches, a maximum of 20 documents per batch is recommended~\cite{mu2024large}. The output is a structured data object that maps identified topics to each question, serving as input for subsequent processing in the robustness assessment and evaluation form creation. We ran the LLM locally using Ollama, a tool for downloading and executing LLMs. Local execution was crucial in our case to prevent any answers from being exposed to Hyperscalers, which might misuse private information. Ollama’s local API supports structured, parsable output and configurable parameters such as seed and temperature. We set a fixed seed and zero temperature fostering reproducibility across runs. All experiments were performed with Llama~3.3~70B, which was reported to be a competitive model at the time of writing, offering a strong balance between performance and hardware efficiency~\cite{OllamaDocumentation}. The experiments were conducted on a local machine equipped with an AMD EPYC 9354P 32-core processor, 188~GB RAM, and two NVIDIA L40S GPUs.

The LLM-based topic modeling pipeline applies zero-shot prompting to batches of survey responses to iteratively extract concise, human-interpretable topic labels. Zero-shot prompting refers to a setting in which the LLM relies solely on its pre-trained knowledge and is guided only by the provided input context, allowing for a lightweight setup. To improve consistency across batches and reduce semantic drift, previously generated topics are provided as additional context, encouraging reuse of existing topics. After processing all batches, a post-processing step merges semantically overlapping topics to ensure a coherent and non-redundant final topic set. All user and system prompts used for topic extraction and merging are inspired by prior work~\cite{mu2024addressing, wang2023prompting, doi2024topic, mu2024large} and are documented in the replication~package~\cite{replication_package}.

%\subsubsection{RQ3: Validation of Topic Modeling Approach}
\subsubsection{Validation of Topic Modeling Approach}
To validate the robustness of our topic modeling approach, we compute metrics to determine whether the extracted insights represent stable underlying patterns rather than artifacts of random variation. Specifically, we compare the lexical overlap and semantic similarity of the extracted topics per question across 30 repeated runs. We compute pairwise Jaccard similarity scores to evaluate the consistency of topic labels and use sentence embeddings with cosine similarity to capture semantic agreement between differing topic sets. For computing embeddings, we use the \textit{SentenceTransformer('all-MiniLM-L6-v2')} model. According to the documentation, \textit{all-MiniLM-L6-v2} is roughly five times faster than their best-performing model \textit{all-mpnet-base-v2}, while still delivering comparable quality~\cite{reimers_sentence_transformers}. This makes it well suited for quantifying the stability and semantic coherence of the generated topics across runs with identical configurations. The validation is performed across all questions, and results are aggregated to provide a comprehensive picture of robustness.

%\subsubsection{RQ4: Evaluation of Topic Quality}
\subsubsection{Evaluation of Topic Quality}
Human evaluation plays a crucial role in topic modeling, particularly when employing an LLM-based approach. In this study, the evaluation was conducted by 23 experts from academia and industry with experience in software engineering and open source ecosystems.
Unlike traditional methods that produce topic representations through keyword lists, LLMs generate directly interpretable topic names. As a result, conventional evaluation metrics for topic modeling, such as coherence, are not suitable~\cite{mu2024addressing, mu2024large}.
Our evaluation framework is specifically designed to assess the quality of the generated topics. To this end, human raters are presented with a series of questions through a self-contained HTML file, which is automatically generated based on the topics identified for each survey question. This HTML file can be displayed without any installation or internet connection, as everything runs locally. This setup streamlines the rating process and minimizes barriers for raters to engage in the evaluation. The user interface of the HTML page is kept simple by incorporating intuitive HTML elements, such as radio buttons, checkboxes, and text input fields.
Building on this structure, each survey question is displayed with its identified topics, for which raters are asked to answer three binary questions assessing the topic quality:

\begin{enumerate}
    \item \textbf{Is this topic interpretable?} An interpretable topic is one that is clear and understandable, representing a valid and coherent concept. For instance, a topic like "dog blue internet" lacks clarity, while a topic like "customer service" is interpretable.
    \item \textbf{Does the topic fit the question?} A topic is considered relevant if it semantically aligns with the question. For example, in response to the question "Why do you like your job?", a topic like "work-life balance" fits, whereas "customer service" does not.
    \item \textbf{Is the topic too specific?} A topic is considered too specific if it includes details that do not help in creating a relevant topic for the survey's objective. For instance, in response to the question "Why do you like your job?", a topic like "talking to Dave during short breaks" might be too narrow. This could instead fall under a broader category such as "great colleagues." While a topic could also be too general, this aspect is not evaluated in the HTML form to prevent cases where the missing information needed to refine the topic is absent from the responses and could not have been identified by the LLM initially.
\end{enumerate}

Since a topic must be interpretable to assess if it fits the question, the second question is only shown if the topic is interpretable, and the third only if it fits the question. A topic meets all quality criteria if the first two questions are answered with "yes" and the third with "no." Afterwards, raters can identify duplicate topics, defined as topics conveying the same meaning and suitable for merging. If no duplicates are identified, the field is left blank. The completed evaluations are then used to assess topic quality across multiple dimensions:

\begin{enumerate}
    \item \textbf{Proportion of topics meeting all three quality criteria:} High values reflect strong topic quality, while low values suggest improvements in the topic modeling method.
    
    \item \textbf{Proportion of uninterpretable topics (not fulfilling the first quality criterion):} High values indicate that raters were unable to understand many topics semantically, suggesting that the LLM produced incoherent topic names.
    
    \item \textbf{Proportion of topics not fitting the question (only fulfilling the first criterion):} High values suggest that the topics fail to semantically match the question, potentially due to LLM hallucinations or irrelevant survey answers. One reason that may lead to irrelevant answers is poorly phrased survey questions.

    \item \textbf{Proportion of too specific topics (only fulfilling the first two criteria):} High values suggest that the topics may be too narrow and could benefit from being more general by combining a subset of the generated topics into a single topic. This may indicate the need for improvements in the topic modeling or topic merging prompts.
    
    \item \textbf{Proportion of duplicate topics:} High values suggest that certain topics may need to be merged, indicating that the topic-merging strategy likely requires refinement.
\end{enumerate}

To quantify the consistency of the expert ratings, we calculated inter-rater agreement for each of the four dimensions: whether a topic is interpretable, whether it fits the question, whether it is too specific, and whether it should be marked as a duplicate. For each dimension, Randolph’s Kappa was computed over all responses at the level of individual questions and in aggregated form across all questions. Randolph’s Kappa is well-suited to our setup as it corrects for chance agreement and considers categorical judgments from multiple raters~\cite{chaturvedi2015evaluation, randolph2005free}.

%% file: sections/4_Results.tex
\section{Results}
As of June 17, 2024, PyPI hosted 547,746 libraries with 186,721 unique email addresses. From these, 50,000 (26.8\%) were randomly selected for survey participation. The first survey on repository URLs, conducted between July 17, 2024 and March 18, 2025, produced 1,385 responses, with 5,722 undeliverable emails, resulting in a 3.1\% response rate---typical for expert-targeted web surveys~\cite{sauermann2013increasing}. For the second one on donation platform URLs, conducted between April 9, 2025 and July 28, 2025, we recontacted 570 participants who had consented to follow-up. Of these, 2 emails bounced and 67 responded, giving a response rate of 11.8\%.

\subsection{RQ1: Topics in Repository URL Survey}
\label{sec:topics_in_repository_url_survey}

To answer \textbf{RQ1}, we included several survey questions to better understand the practices of package maintainers regarding source code repository links. Each question targets a different aspect, such as platform preferences, link maintenance, or the reasons for not assigning repository URLs. Table~\ref{tab:topics_repository_url} summarizes the topics identified from the answers to these questions (\textbf{SQ-1.1}: 451 responses, \textbf{SQ-1.2}: 792 responses, \textbf{SQ-1.3}: 456 responses).
% Repository
%   1.1 -> 5
%   1.2 -> 1
%   1.3 -> 3

\begin{table}[ht]
\small
\centering
\caption{Identified topics in the repository URL survey}
\setlength{\tabcolsep}{6pt}
\resizebox{\tableWidth\textwidth}{!}{%
\begin{tabular}{p{0.48\linewidth} | p{0.48\linewidth}}
\hline
\textbf{Survey Question} & \textbf{Discovered Topics} \\
\hline

\textbf{SQ-1.1:} If you assigned a GitLab, Bitbucket, Gitea, Codeberg or sourcehut URL to your PyPI library, what were the reasons for not hosting the code on GitHub and linking it accordingly?
%(5)
&
Avoiding Proprietary; Decentralized Tech; Feature Preference; Diversifying Platforms; Dislike GitHub; Institutional Choice. \\
\hline

\textbf{SQ-1.2:} Can you please (double) check that your assigned URL is still working. If it returns an error code 404, why haven't you updated the URL accordingly?
%(1)
&
Still Working; URL Issues; Project Status; Link Functionality. \\
\hline

\textbf{SQ-1.3:} If you have not assigned any URL to your PyPI library, what were the reasons for this?
%(3)
&
Forgot to Assign; Lack Awareness Of Process; Not Serious Project; No Maintenance Needed; Template Default Used; No Advantage Perceived; No GitHub Repository; Not Applicable To Project; Lazy To Assign; No Reason Provided; Not Publicly Available; No Need To Assign. \\

\hline
\end{tabular}
}
\label{tab:topics_repository_url}
\end{table}

In \textbf{SQ-1.1}, which investigated the reasons for using alternatives to GitHub (such as GitLab or Bitbucket), respondents articulated values related to \textit{ideological stance}, \textit{technical preference}, and \textit{organizational constraints}. The topics \textit{Avoiding Proprietary}, \textit{Decentralized Tech}, and \textit{Dislike GitHub} suggest ideological or ethical objections to GitHub or centralized platforms in general (e.g., \textit{``I stopped using GitHub when Microsoft bought it''}, \textit{``No-trust, in Microsoft which owns GitHub''}, \textit{``GitHub is a proprietary platform owned by Microsoft''}, \textit{``Monopolies suck''}, \textit{``too much centralization at one provider is not good''}). In contrast, \textit{Feature Preference} and \textit{Diversifying Platforms} point to a more pragmatic rationale based on functionality or risk mitigation (e.g., \textit{``[...] GitLab had CI features and GitHub did not''}, \textit{``[...] Bitbucket supported private repos for free''}, \textit{``I wanted to promote diversity of the code hosting ecosystem''}). \textit{Institutional Choice} reflects scenarios where the platform is selected at the organizational level, beyond the control of individual developers (e.g., \textit{``gitlab instance maintained by my scientific collaboration''}, \textit{``Our company uses gitlab [...]''}).

For \textbf{SQ-1.2}, which asked about outdated URLs, the identified topics revolve around \textit{maintenance status} and \textit{awareness}. Responses indicating \textit{Still Working} or \textit{Link Functionality} show that many participants verified that their links remained valid (e.g., \textit{``It still works''}, \textit{``All of my links are working''}). Others referenced \textit{URL Issues} as reasons for not updating the repository links, often linked to lack of awareness or prioritization (e.g., \textit{``did not know that they were broken''}, \textit{``If there is 404, then it hasn't been brought up to my attention''}, \textit{``Most probably forgot''}, \textit{``Old projects so did not care much''}). Additionally, several responses highlight practical barriers such as missing automation or time constraints (e.g., \textit{``there doesn't seem to be an automated tool to remind me''}, \textit{``no automation in place for checking this''}, \textit{``Not enough time''}), as well as cases related to \textit{Project Status}, where projects are abandoned or no longer actively maintained (e.g., \textit{``The project aren't maintained anymore''}, \textit{``No active development''}), suggesting that updating metadata is often deprioritized for inactive projects.

Finally, \textbf{SQ-1.3} explored why projects lacked any assigned URL. Here, the responses were diverse but grouped primarily into themes of \textit{oversight}, \textit{project nature}, and \textit{lack of perceived value}. Some developers indicated forgetfulness or lack of awareness (\textit{Forgot To Assign}, \textit{Lack Awareness Of Process}), for example \textit{``Because I forgot''}, \textit{``Forgot to include it in the pyproject.toml''}, or \textit{``I wasn't aware how to do it''}. Others suggested their project was not mature or intended for serious use (\textit{Not Serious Project}, \textit{No Maintenance Needed}, \textit{Template Default Used}), e.g., \textit{``The project is not serious enough to warrant one''}, \textit{``My PyPI project never matured beyond the "alpha test" stage''}, or \textit{``Mistakenly forgot to do it. Often I use a project template to start projects, then the field is normally already filled in''}. Several topics also reflect perceived irrelevance or lack of necessity (\textit{No Advantage Perceived}, \textit{Not Applicable To Project}, \textit{No Need To Assign}), for instance \textit{``Didn't seem necessary for a small project''} or \textit{``If the library is not something in active development - who cares?''}. Additionally, barriers such as the absence of a public repository or intentional restriction of access were mentioned (\textit{No GitHub Repository}, \textit{Not Publicly Available}), e.g., \textit{``Did not have a public URL to assign''} or \textit{``The github repos are private''}. Some also cited motivational factors such as laziness (e.g., \textit{``Laziness''}, \textit{``Too lazy''}) or did not provide a concrete reason (e.g., \textit{``No Reason''}).

\subsection{RQ2: Topics in Donation Platform URL Survey}
\label{sec:topics_in_donation_platform_url_survey}

To answer \textbf{RQ2}, Table~\ref{tab:topics_donation_platform_url} summarizes maintainers' practices regarding donation platform URLs. Each survey question addresses a different aspect, including platform dominance, link maintenance, and the reasons for not assigning donation platform links, based on 10 responses for \textbf{SQ-2.1}, 6 for \textbf{SQ-2.2}, and 61 for \textbf{SQ-2.3}. Despite the modest number of responses compared to the repository URL survey, the results still provide diverse and valuable insights.

% Donation platform
%   2.1 -> 8
%   2.2 -> 4
%   2.3 -> 1

\begin{table}[ht]
\small
\centering
\caption{Identified topics in donation platform URL survey}
\setlength{\tabcolsep}{6pt}
\resizebox{\tableWidth\textwidth}{!}{%
\begin{tabular}{p{0.48\linewidth} | p{0.48\linewidth}}
\hline
\textbf{Survey Question} & \textbf{Discovered Topics} \\
\hline

\textbf{SQ-2.1:} If you assigned a link to any donation platform exclusively on the GitHub repository, but not on the library's PyPI project page, what were the reasons for this? 
%(8)
& 
Lack Of Knowledge; Oversight Error; Visibility Preference; Referral Strategy; Technical Difficulty. \\
\hline

\textbf{SQ-2.2:} If you have linked a GitHub Sponsors profile on your library's PyPI project page, but the URL is not reachable (status code 404) anymore, what are the reasons for this? 
%(4)
& 
Not Applicable; No Link; URL Shift. \\
\hline

\textbf{SQ-2.3:} If you have not assigned a link to any donation platform on the library's PyPI project page or on the GitHub repository, what were the reasons for this? 
%(1)
& 
No Need; Job Compensation; Unaware Options; Personal Projects; Public Benefit; No Profit; Organizational Policy; Skepticism; No Account. \\

\hline
\end{tabular}
}
\label{tab:topics_donation_platform_url}
\end{table}

\textbf{SQ-2.1} explored why donation links appeared preferentially on GitHub repositories and not on PyPI. Here, reasons were tied to \textit{usability constraints} and \textit{knowledge gaps}. Topics like \textit{Lack Of Knowledge}, \textit{Oversight Error}, and \textit{Technical Difficulty} highlight friction in the process (e.g., \textit{``Forgot''}, \textit{``An oversight''}, \textit{``I can't figure out how to setup the link in PyPI. It's not obvious.''}, \textit{``I did not know that you can link to these in the pyproject.toml file''}). In contrast, \textit{Visibility Preference} and \textit{Referral Strategy} suggest strategic placement decisions for better exposure or timing (e.g., \textit{``Better visibility''}, \textit{``no one will visit the PyPI project page, right?''}, \textit{``At the time someone is looking at PyPI, they're likely doing software selection, and therefore it's too early to start asking things of them''}).

\textbf{SQ-2.2} looked at outdated GitHub Sponsors URLs. The limited set of responses revealed only a small number of practical issues. Some responses pointed to potential link instability, for instance due to missing monitoring (e.g., \textit{``No automatic URL checking means sometimes the URL will shift under us''}). Others responded that no donation link was provided or that the situation did not apply (e.g., \textit{``Nothing linked''}, \textit{``I have not encountered this issue yet''}).

Finally, \textbf{SQ-2.3} addresses why no donation links were included at all. The identified topics point to a mix of \textit{personal}, \textit{practical}, and \textit{ideological barriers}. Some developers expressed a lack of need due to personal circumstances (\textit{No Need}, \textit{Job Compensation}, \textit{No Profit}, \textit{Personal Projects}), for example \textit{``No need for donations''}, \textit{``Maintaining my PyPI projects is part of my job, for which I already get paid''}, \textit{``I have no intention of making any profit on my PyPI project''}, and \textit{``My work is purely a hobby''}. Others highlighted practical constraints such as \textit{Unaware Options}, \textit{No Account}, and \textit{Administrative Burden}, including \textit{``I wasn't aware of the possibility''}, \textit{``I do not have an account on a donation platform''}, and \textit{``administration burden''}. Aspects regarding \textit{Organizational Policy} were also mentioned, such as \textit{``Most projects are part of a larger organization and they should set a policy for this''} and \textit{``library belong to public institution''}. Finally, some cited ideological positions such as \textit{Skepticism} and \textit{Public Benefit}, where monetization is questioned or non-financial motivations are preferred (e.g., \textit{``I code for fun, not for money''}, \textit{``My open source contributions are for the benefit of the public''}).

%\subsection{RQ3: Robustness of Topic Modeling Approach}
\subsection{Robustness of Topic Modeling Approach}
\label{sec:results_robustness}

%To answer \textbf{RQ3}, we evaluate how consistent the extracted topics are across all repeated runs.

To validate the robustness of our topic modeling approach, we assess the consistency of the extracted topics across all repeated runs. Table~\ref{tab:robustness} reports the lexical and semantic similarity of the identified topics across 30 runs per question, using Jaccard and Cosine similarity as robustness metrics.
The results indicate that our topic modeling approach demonstrates a high degree of robustness. On average, the Jaccard and Cosine similarity scores exceed 0.8 for each survey, indicating substantial lexical and semantic consistency in the extracted topics. As expected, Cosine similarity is consistently equal to or higher than Jaccard similarity, reflecting cases where topic sets differ slightly in wording but remain semantically equivalent. Some variability is observed for specific questions (e.g., \textbf{SQ-1.3} and \textbf{SQ-2.2}), which may be due to lower data quality in the corresponding survey responses. Nonetheless, most questions yield near-perfect similarity scores, underlining the reliability of our topic modeling approach.

\begin{table}[ht]
\small
\centering
\caption{Robustness of topic modeling approach}
\resizebox{\textwidth}{!}{%
\begin{tabular}{c c c c c}
\hline
\textbf{Survey} & \makecell{\textbf{Survey Question}} & \makecell{\textbf{Number Runs}} & \makecell{\textbf{Jaccard Similarity}} & \makecell{\textbf{Cosine Similarity}} \\
\hline
\multirow{3}{*}{\makecell{\textbf{Repository} \\ \textbf{URL}}} 
& 1.1 & 30 & 1.0 & 1.0 \\
& 1.2 & 30 & 0.96 & 0.97 \\
& 1.3 & 30 & 0.57 & 0.66 \\
\cline{1-5}
& \textbf{Average} & \textbf{30} & \textbf{0.84} & \textbf{0.88} \\
\hline
\multirow{3}{*}{\makecell{\textbf{Donation} \\ \textbf{Platform} \\ \textbf{URL}}} 
& 2.1 & 30 & 1.0 & 1.0 \\
& 2.2 & 30 & 0.75 & 0.89 \\
& 2.3 & 30 & 1.0 & 1.0 \\
\cline{1-5}
& \textbf{Average} & \textbf{30} & \textbf{0.92} & \textbf{0.96} \\
\hline
%\textbf{Overall} & \textbf{Average} & \textbf{30} & \textbf{0.84} & \textbf{0.89} \\
%\hline
\end{tabular}
}
\label{tab:robustness}
\end{table}

%\subsection{RQ4: Human Evaluation of Topic Quality}
\subsection{Human Evaluation of Topic Quality}
\label{sec:results_human_evaluation}

%To answer \textbf{RQ4}, \numRaters experts from industry and research assessed the quality of the identified topics for each survey question across the two surveys. 
We conducted a human evaluation with \numRaters experts from industry and research, who assessed the quality of the identified topics for each survey question across the two surveys. Table~\ref{tab:eval} summarizes the outcomes measured by the number of evaluated topics, the proportion meeting all quality criteria, and the proportions of topics flagged as uninterpretable, irrelevant, too specific, or duplicates. We also quantify the inter-rater agreement using Randolph’s Kappa.

\begin{table}[ht]
\small
\centering
\caption{Human evaluation results assessing topic quality among \numRaters\ raters}
\setlength{\tabcolsep}{3pt}
\resizebox{\textwidth}{!}{%
\begin{tabular}{l | c c | c c c c c | c}
\hline
\textbf{Survey} & \makecell{\textbf{Survey}\\\textbf{Question}} &
\makecell{\textbf{Number}\\\textbf{Topics}} &
\makecell{\textbf{Topics}\\\textbf{Meeting}\\\textbf{Quality}\\\textbf{Criteria}} &
\makecell{\textbf{Uninterpretable}\\\textbf{Topics}} &
\makecell{\textbf{Topics}\\\textbf{Not}\\\textbf{Fitting}\\\textbf{Question}} &
\makecell{\textbf{Topics}\\\textbf{Too}\\\textbf{Specific}} &
\makecell{\textbf{Duplicate}\\\textbf{Topics}} &
\makecell{\textbf{Randolph's}\\\textbf{Kappa}} \\
\hline

\multirow{3}{*}{\makecell{\textbf{Repository}\\\textbf{URL}}}
& 1.1 & 6 & 0.83 & 0.09 & 0.04 & 0.03 & 0.17 & 0.61 \\ % 1.1 -> 5
& 1.2 & 4 & 0.59 & 0.20 & 0.16 & 0.05 & 0.32 & 0.40 \\ % 1.2 -> 1
& 1.3 & 12 & 0.81 & 0.06 & 0.06 & 0.07 & 0.24 & 0.60 \\ % 1.3 -> 3
\cline{1-9}
& \textbf{Overall} & \textbf{22} & \textbf{0.78} & \textbf{0.09} & \textbf{0.08} & \textbf{0.06} & \textbf{0.24} & \textbf{0.57} \\
\hline

\multirow{3}{*}{\makecell{\textbf{Donation}\\\textbf{Platform}\\\textbf{URL}}}
& 2.1 & 5 & 0.83 & 0.10 & 0.04 & 0.03 & 0.17 & 0.62 \\ % 2.1 -> 8
& 2.2 & 3 & 0.70 & 0.19 & 0.12 & 0.00 & 0.26 & 0.40 \\ % 2.2 -> 4
& 2.3 & 9 & 0.77 & 0.12 & 0.06 & 0.05 & 0.22 & 0.55 \\ % 2.3 -> 1
\cline{1-9}
& \textbf{Overall} & \textbf{17} & \textbf{0.77} & \textbf{0.13} & \textbf{0.06} & \textbf{0.04} & \textbf{0.21} & \textbf{0.54} \\
\hline
\end{tabular}
}
\label{tab:eval}
\end{table}

For the \textbf{repository URL survey}, a total of 22 topics were evaluated. On average, 78\% of these topics fulfilled all three quality criteria (\textit{interpretable}, \textit{fitting the question}, \textit{not too specific}), while only 9\% were judged as uninterpretable and 8\% as irrelevant. A small fraction (6\%) were marked as too specific, whereas nearly one quarter (24\%) were identified as duplicates. Inter-rater agreement varied across the four evaluation dimensions. Note that only the aggregated Randolph’s Kappa is reported in Table~\ref{tab:eval}. Agreement was highest for \textit{is\_interpretable} ($\kappa = 0.78$, substantial agreement), followed by \textit{is\_fitting\_question} ($\kappa = 0.61$, substantial agreement) and \textit{is\_too\_specific} ($\kappa = 0.52$, moderate agreement). In contrast, agreement on duplicates was notably lower ($\kappa = 0.34$, fair agreement). The overall average for the repository URL survey is $\kappa = 0.57$, which falls into the \textit{moderate agreement} range~\cite{yilmaz2017assessing}.

For the \textbf{donation platform URL survey}, raters assessed 17 topics. Here, 77\% of topics met all quality criteria, similar to the repository URL survey. A slightly higher proportion (13\%) were considered as uninterpretable, while 6\% as irrelevant, and only 4\% as too specific. Duplicate topics accounted for 21\%, consistent with the repository survey. Inter-rater agreement also showed variations. Again, only the aggregated Randolph’s Kappa is reported in Table~\ref{tab:eval}. Raters reached substantial agreement on \textit{is\_interpretable} ($\kappa = 0.68$), while agreement was moderate for \textit{is\_fitting\_question} ($\kappa = 0.54$), \textit{is\_too\_specific} ($\kappa = 0.50$) and duplicate topics ($\kappa = 0.44$). The overall average Randolph’s Kappa for the donation platform URL survey was $\kappa = 0.54$, also indicating \textit{moderate agreement}~\cite{yilmaz2017assessing}.

Taken together, these results show that the majority of discovered topics were rated as high quality, with only a minority presenting interpretability or relevance issues. Agreement among raters was consistently stronger on aspects of interpretability and relevance, while assessments of specificity and duplicate detection were more subjective and less consistent. Both surveys achieved moderate inter-rater agreement, underlining the reliability of the evaluation, while certain judgments---such as whether a topic is too specific or a duplicate---involve subjective interpretation, leading to variation among raters. In addition, variation across questions are observed for all evaluation metrics, which may reflect limitations of the topic modeling pipeline (e.g., improve topic merging stage) or survey design and data quality issues, such as ambiguous survey questions leading to low-quality responses.

%% file: sections/5_Discussion.tex
\section{Discussion}

\subsection{Understanding Metadata as a Socio-Technical Problem}

Our findings suggest that incomplete or outdated metadata in PyPI is not merely a technical issue, but results from a combination of maintainer behavior (e.g., oversight, lack of awareness, or low perceived importance), platform limitations (e.g., missing guidance and validation), and project-specific factors (e.g., project maturity and maintenance status). These findings indicate that maintainers often do not treat metadata as a first-class artifact, but rather as optional or secondary to code development. At the same time, decisions around platform usage (e.g., GitHub dominance) and donation link placement reflect deeper ideological, organizational, and strategic considerations. Maintainers balance convenience, visibility, and personal or institutional values when deciding how and where to expose project metadata. This reinforces the view that metadata practices cannot be improved solely through technical mechanisms, but require interventions that adopt a more human-centered perspective, focusing on the behavior and needs of maintainers. This aligns with recent work emphasizing the socio-technical nature of OSS ecosystems~\cite{feng2026charting}.

\subsection{Implications for Package Registry Design}

A key implication of our findings is that current package registry interfaces, including PyPI, may not sufficiently support maintainers in providing complete and up-to-date metadata. The frequent occurrence of topics such as \textit{Forgot To Assign}, \textit{Lack Awareness Of Process}, and \textit{URL Issues} points to usability gaps and missing feedback mechanisms. Improving metadata quality therefore requires rethinking registry design along two key dimensions:
First, maintainers often lack awareness of available metadata fields or their importance, suggesting the need for improved guidance and onboarding. Integrating contextual support directly into the publishing workflow---such as examples, tooltips, or best-practice templates---could lower the barrier to providing complete metadata. Second, the absence of automated checks allows missing or outdated links to persist unnoticed, indicating the need for better validation and feedback mechanisms. Lightweight approaches, such as warnings for missing repository URLs or periodic checks for broken links, could significantly improve metadata reliability without imposing substantial overhead. This is consistent with prior proposals, which advocate automated validation of repository metadata and highlighting the need for improved linkage between package registries and source repositories~\cite{gao2024pyradar, tsakpinis_pretschner_2024}.

\subsection{Rethinking Visibility and Incentives for Donation Links}

The results of the donation platform survey highlight a misalignment between where funding information is provided and where it could be most effective. Maintainers frequently prefer GitHub over PyPI for displaying donation links due to perceived higher visibility and better integration with existing workflows. This behavior suggests that PyPI is not currently perceived as a suitable location for funding-related information. The reasons are twofold: first, maintainers often lack awareness of how to add such links; second, they question the value of doing so in a context primarily associated with package distribution rather than community engagement. To address this, package registries could play a more active role in normalizing and promoting funding metadata. Possible approaches include increasing the prominence of donation links within project pages and providing standardized mechanisms aligned with widely adopted formats such as \texttt{FUNDING.yml}. This could involve enabling automatic synchronization of donation platform information from source code repositories, as well as communicating the benefits of funding visibility for project sustainability. However, it is equally important to recognize that not all maintainers are motivated by financial incentives. Topics such as \textit{No Need}, \textit{Public Benefit}, and \textit{Skepticism} indicate that donation links are influenced by personal values and project context. As such, any platform-level intervention should remain optional and avoid framing funding as a universal expectation.

\subsection{Cross-Cutting Challenges in Metadata Maintenance}

Across both repository and donation link practices, a consistent pattern emerges: metadata degradation is often gradual and unintentional. Outdated links, missing entries, and inconsistent placement are rarely the result of deliberate decisions, but rather of limited attention, competing priorities, and lack of tooling support. This highlights the need for continuous rather than one-time interventions. Metadata quality cannot be ensured solely at publication time, but instead requires ongoing maintenance. Mechanisms such as periodic notifications, for example via email to maintainers~\cite{pypi_email_issue}, dashboards highlighting incomplete metadata, or ecosystem-level monitoring tools could help maintainers keep metadata up to date.

\subsection{Implications for Software Supply Chain Security and Sustainability}

Metadata completeness has direct implications for security and sustainability in OSS ecosystems. Missing or outdated repository links hinder the ability to monitor maintenance activity, detect vulnerabilities, and assess project health. Similarly, the absence of donation links reduces the visibility of funding opportunities, potentially limiting the long-term sustainability of projects. Improving metadata quality can enhance dependency analysis, facilitate automated security assessments, and support funding mechanisms---all of which contribute to a more resilient software supply chain. From this perspective, package registries should be viewed not only as distribution platforms but also as infrastructure that enables better monitoring of dependencies, maintenance activities, and funding opportunities.

\subsection{Methodological Reflections on LLM-Based Topic Modeling}

This study provides empirical insights into LLM-based topic modeling for analyzing large-scale survey data in a short-text settings, complementing recent work by highlighting strengths and limitations~\cite{wang2023prompting, doi2024topic}. Our robustness analysis indicates that the approach produces stable results across repeated runs, with high lexical and semantic similarity. This suggests that LLM-based topic modeling can reliably capture dominant themes in short-text datasets. At the same time, variability observed in questions with noisier input highlights the dependency on data quality and survey design. The human evaluation further shows that most generated topics are interpretable and relevant, while challenges remain in controlling granularity and avoiding duplicate topics. These findings suggest that LLM-based topic modeling can be a viable and scalable alternative to manual qualitative analysis. However, it benefits from complementary human validation, especially when fine-grained distinctions between topics are required. Future work could explore hybrid approaches that combine LLM-based topic generation with human-assisted post-processing techniques for improved topic consolidation, as well as more advanced preprocessing methods such as semantic filtering to reduce noise in input data.
A key limitation of our approach is that direct evaluation of topics against the underlying survey responses is not possible. This is due to the generative nature of the method, which processes documents in batches of 10 and does not explicitly link individual responses to the generated topics. Even if such a mapping were available, evaluating topics against individual responses would be infeasible at scale due to the required annotation effort. Instead, assessing whether topics are interpretable and semantically aligned with the survey questions provides a practical form of indirect validation and remains feasible even at scale.

%% file: sections/6_Threats_to_Validity.tex
\section{Threats to Validity}
Threats to validity are divided into the following four aspects~\cite{runeson2009guidelines}: 

\textbf{Construct validity:}
%describes \say{to what extent the operational measures that are studied really represent what the researcher has in mind and what is investigated according to the research questions}~\cite{runeson2009guidelines}. 
There is a potential threat to construct validity concerning the input dataset used for the topic modeling approach. The dataset is derived from responses to two surveys, whose questions were designed based on notable empirical findings from related work~\cite{tsakpinis_pretschner_2024, tsakpinis2025analyzing}. A threat may arise if the questions were not precisely formulated to uncover the reasons behind these notable findings or if they were poorly worded, both leading to confusion and unreliable responses. To mitigate this risk, we ensured that our survey questions were clearly formulated to minimize potential misinterpretation and were strictly based on the empirical findings from previous research~\cite{tsakpinis_pretschner_2024, tsakpinis2025analyzing}. A further threat arises from the recruitment strategy and the lack of participant information. As the survey was distributed via email using publicly available contact data, the respondent population could not be controlled, which may have resulted in responses of varying quality. In addition, we did not collect demographic or expertise-related information to preserve anonymity and reduce survey friction, limiting our ability to assess potential biases.

\textbf{Internal validity:} 
%\say{is of concern when causal relations are examined}~\cite{runeson2009guidelines}. 
This study does not aim to establish causal relationships but to explore factors associated with notable metadata practices. Preprocessing decisions and the LLM-based topic modeling approach (e.g., prompt design and batching) may have influenced the results. We mitigate these threats through repeated runs with fixed configurations as part of the robustness analysis and through complementary human~evaluation.

\textbf{External validity:} 
%is concerned with \say{to what extent it is possible to generalize the findings, and to what extent the findings are of interest to other people outside the investigated case}~\cite{runeson2009guidelines}. 
We focus on PyPI as our ecosystem of study. While our findings may not directly generalize to other ecosystems, the methodology is applicable to any publicly accessible ecosystem where libraries include links (e.g., repositories or donation platforms) in their metadata and contain contact information about their authors and maintainers.

\textbf{Reliability:} 
%refers \say{to what extent the data and the analysis are dependent on the specific researchers}~\cite{runeson2009guidelines}.
A threat to reliability stems from the conflict between ensuring reproducibility and protecting the privacy of survey participants. While participants' email addresses and the survey answers are excluded from the replication package to prevent any leakage of personal data~\cite{verdecchia2025notes}, the topic modeling pipeline, its results and the evaluation data is available on Figshare~\cite{replication_package}. Despite using a fixed seed and a temperature of zero when prompting the LLM, a minor threat remains, as the model still exhibits slight deviations in results across repeated runs with identical configurations---as indicated by our robustness evaluations.

%% file: sections/7_Conclusion_and_Future_Work.tex
\section{Conclusion and Future Work}

This paper investigated the factors associated with notable metadata practices in PyPI libraries, specifically missing, outdated, and platform-dominated links to code repositories and donation platforms. Based on 1,776 open-ended survey responses, analyzed using an LLM-based topic modeling pipeline, we identified a diverse set of factors characterizing these practices. Our findings show that repository links are often missing due to oversight, lack of awareness, or perceived irrelevance, while platform dominance is shaped by ideological, technical, and organizational considerations. Similarly, donation platform links remain sparse due to skepticism, limited perceived benefit, and usability barriers, and are frequently placed on GitHub rather than PyPI for visibility reasons. Taken together, these findings highlight concrete opportunities to improve PyPI’s metadata practices. In particular, clearer guidance during package publication, improved usability of metadata interfaces, and automated checks for outdated or missing links could help increase metadata completeness. This, in turn, can strengthen security assessments by enabling better monitoring of maintenance activities, while also supporting OSS sustainability through increased visibility of funding options. Beyond these empirical insights, we demonstrated that LLM-based topic modeling using LLaMA~3.3~70B can serve as a reliable and scalable approach for analyzing large-scale qualitative data in software engineering. Our evaluation shows high robustness measured by similarity metrics (Jaccard 0.84, cosine 0.92) and generally high-quality topics (77--78\% meeting all criteria), with moderate agreement among human raters ($\kappa \approx 0.55$), supporting the use of such methods as practical alternative to manual qualitative analysis.

Future work should employ purposive or stratified sampling and more targeted participant selection strategies to capture more diverse maintainer profiles and obtain deeper insights into specific maintainer groups~\cite{baltes2022sampling}. In addition, investigating semantic filtering techniques offers a promising direction for reducing noise in the input dataset. Finally, research could examine platform-level interventions on PyPI to assess how they influence authors’ and maintainers’ behavior in adding and updating repository and donation platform links.

%% file: sections/8_Data_Availability.tex
\section{Data Availability}
\label{sec:data_availabaility}
This study did not require formal ethics approval, as it involved voluntary, anonymized survey responses and did not collect sensitive personal data. To protect the privacy of participants, individual survey responses and email addresses cannot be shared publicly. All other artifacts required to reproduce the results are available in our replication package on Figshare~\cite{replication_package}. This includes: (1) both surveys, (2) datasets and scripts required to reproduce the quantitative results from related work, (3) the LLM-based topic modeling pipeline including prompts and configuration, and (4) the evaluation framework, including HTML forms and collected expert ratings. Researchers seeking access to additional data may contact the corresponding author. Access may be granted subject to ethical considerations and a data use agreement.

%% file: research_paper.bib
@inproceedings{tsakpinis2023analyzing,
  title={Analyzing Maintenance Activities of Software Libraries},
  author={Tsakpinis, Alexandros},
  booktitle={Proceedings of the 27th International Conference on Evaluation and Assessment in Software Engineering},
  pages={313--318},
  year={2023},
  doi          = {10.1145/3593434.3593474}
}

@article{ebert2008open,
  title={Open source software in industry},
  author={Ebert, Christof},
  journal={IEEE Software},
  volume={25},
  number={3},
  pages={52--53},
  year={2008},
  publisher={IEEE},
  doi          = {10.1109/MS.2008.67}
}

@article{pittenger2016open,
  title={Open source security analysis: The state of open source security in commercial applications},
  author={Pittenger, Mike},
  journal={Black Duck Software, Tech. Rep},
  year={2016}
}

@inproceedings{bauer2012structured,
  title={A structured approach to assess third-party library usage},
  author={Bauer, Veronika and Heinemann, Lars and Deissenboeck, Florian},
  booktitle={2012 28th IEEE International Conference on Software Maintenance (ICSM)},
  pages={483--492},
  year={2012},
  organization={IEEE},
  doi={10.1109/ICSM.2012.6405311},
}

@article{cox2019surviving,
  title={Surviving software dependencies},
  author={Cox, Russ},
  journal={Communications of the ACM},
  volume={62},
  number={9},
  pages={36--43},
  year={2019},
  publisher={ACM New York, NY, USA},
  doi={10.1145/3347446}
}

@inproceedings{decan2018impact,
  title={On the impact of security vulnerabilities in the npm package dependency network},
  author={Decan, Alexandre and Mens, Tom and Constantinou, Eleni},
  booktitle={Proceedings of the 15th international conference on mining software repositories},
  pages={181--191},
  year={2018},
  doi={10.1145/3196398.3196401}
}

@inproceedings{rahkema2022swiftdependencychecker,
  title={SwiftDependencyChecker: Detecting Vulnerable Dependencies Declared Through CocoaPods, Carthage and Swift PM},
  author={Rahkema, Kristiina and Pfahl, Dietmar},
  booktitle={9th International Conference on Mobile Software Engineering and Systems (MobileSoft)},
  pages={107--111},
  year={2022},
  organization={IEEE},
  doi          = {10.1145/3524613.3527806}
}

@inproceedings{raemaekers2011exploring,
  title={Exploring risks in the usage of third-party libraries},
  author={Raemaekers, Steven and van Deursen, Arie and Visser, Joost},
  booktitle={BElgian-NEtherlands software eVOLution seminar},
  pages={31},
  year={2011}
}

@article{runeson2009guidelines,
  title={Guidelines for conducting and reporting case study research in software engineering},
  author={Runeson, Per and H{\"o}st, Martin},
  journal={Empirical software engineering},
  volume={14},
  number={2},
  pages={131--164},
  year={2009},
  publisher={Springer},
  doi          = {10.1007/S10664-008-9102-8}
}

@article{abdalkareem2020impact,
  title={On the impact of using trivial packages: An empirical case study on npm and pypi},
  author={Abdalkareem, Rabe and Oda, Vinicius and Mujahid, Suhaib and Shihab, Emad},
  journal={Empirical Software Engineering},
  volume={25},
  pages={1168--1204},
  year={2020},
  publisher={Springer},
  doi={10.1007/S10664-019-09792-9},
}

@misc{Octoverse2025,
  author = {GitHub},
  title  = {Octoverse: A new developer joins GitHub every second as AI leads TypeScript to \#1},
  year   = {2025},
  url    = {https://github.blog/news-insights/octoverse}
}

@article{bommarito2019empirical,
  title={An Empirical Analysis of the Python Package Index (PyPI)},
  author={Bommarito, Ethan and Bommarito II, Michael J},
  journal={arXiv preprint},
  year={2019},
  doi = {10.48550/arXiv.1907.11073},
}

@inproceedings{decan2016topology,
  title={On the topology of package dependency networks: A comparison of three programming language ecosystems},
  author={Decan, Alexandre and Mens, Tom and Claes, Maelick},
  booktitle={Proccedings of the 10th european conference on software architecture workshops},
  pages={1--4},
  year={2016},
  doi={10.1145/2993412.3003382}
}

@misc{replication_package,
  author = {Anonymous},
  title  = {Replication Package},
  year   = {2026},
  url    = {https://figshare.com/s/b510f5274eb1333deb95},
  note   = {Figshare dataset}
}

@inproceedings{tsakpinis_pretschner_2024,
author = {Tsakpinis, Alexandros and Pretschner, Alexander},
title = {Analyzing the Accessibility of GitHub Repositories for PyPI and NPM Libraries},
booktitle = {Proceedings of the 28th International Conference on Evaluation and Assessment in Software Engineering},
year = {2024},
pages = {345--350},
doi          = {10.1145/3661167.3661231},
}

@inproceedings{tsakpinis2025analyzing,
  title={Analyzing the Usage of Donation Platforms for PyPI Libraries},
  author={Tsakpinis, Alexandros and Pretschner, Alexander},
  booktitle={Proceedings of the 29th International Conference on Evaluation and Assessment in Software Engineering},
  pages={628--633},
  year={2025},
  doi          = {10.1145/3756681.3757018}
}

@article{chung2022text,
  title={Text-mining open-ended survey responses using structural topic modeling: A practical demonstration to understand parents’ coping methods during the COVID-19 pandemic in Singapore},
  author={Chung, Gerard and Rodriguez, Maria and Lanier, Paul and Gibbs, Daniel},
  journal={Journal of Technology in Human Services},
  volume={40},
  number={4},
  pages={296--318},
  year={2022},
  publisher={Taylor \& Francis},
  doi = {10.31219/osf.io/enzst},
}

@article{pietsch2018topic,
  title={Topic modeling for analyzing open-ended survey responses},
  author={Pietsch, Andra-Selina and Lessmann, Stefan},
  journal={Journal of Business Analytics},
  volume={1},
  number={2},
  pages={93--116},
  year={2018},
  publisher={Taylor \& Francis},
  doi={10.1080/2573234X.2019.1590131}
}

@misc{oss2022,
  author = {{OpenSSF}},
  title  = {The Open Source Software Security Mobilization Plan},
  year   = {2022},
  url    = {https://openssf.org/oss-security-mobilization-plan/}
}

@misc{tidelift2024,
  author = {Tidelift},
  title  = {The 2024 Tidelift State of the Open Source Maintainer Report},
  year   = {2024},
  url    = {https://tidelift.com/open-source-maintainer-survey-2024}
}

@article{medappa2023sponsorship,
  title={Sponsorship Funding in Open-Source Software: Effort Reallocation and Spillover Effects in Knowledge-Sharing Ecosystems},
  author={Medappa, Poonacha K and Tunc, Murat M and Li, Xitong},
  journal={Available at SSRN 4484403},
  year={2023},
  doi={10.2139/ssrn.4484403},
}

@article{hahn2024improving,
  title={Improving and analyzing open-ended survey responses: A case study linking psychological theories and analysis approaches for text data.},
  author={Hahn, Sonja and Kroehne, Ulf and Merk, Samuel},
  journal={Zeitschrift f{\"u}r Psychologie},
  volume={232},
  number={3},
  pages={171},
  year={2024},
  publisher={Hogrefe Publishing},
  doi          = {10.1027/2151-2604/a000566}
}

@article{churchill2022evolution,
  title={The evolution of topic modeling},
  author={Churchill, Rob and Singh, Lisa},
  journal={ACM Computing Surveys},
  volume={54},
  number={10s},
  pages={1--35},
  year={2022},
  publisher={ACM New York, NY},
  doi          = {10.1145/3507900}
}

@article{roberts2014structural,
  title={Structural topic models for open-ended survey responses},
  author={Roberts, Margaret E and Stewart, Brandon M and Tingley, Dustin and Lucas, Christopher and Leder-Luis, Jetson and Gadarian, Shana Kushner and Albertson, Bethany and Rand, David G},
  journal={American journal of political science},
  volume={58},
  number={4},
  pages={1064--1082},
  year={2014},
  publisher={Wiley Online Library},
  doi={10.1111/ajps.12103}
}

@misc{OllamaDocumentation,
  title = {Ollama Documentation},
  author = {{Ollama}},
  year = {2025},
  url = {https://docs.ollama.com/}
}

@misc{pypi-simple,
  title        = {{PyPI Simple API}},
  author       = {{Python Software Foundation}},
  url          = {https://pypi.org/simple/},
  year         = {2026}
}

@misc{pypi-json,
  title        = {{PyPI API}},
  author       = {{Python Software Foundation}},
  url          = {https://pypi.org/pypi/<name>/json},
  year         = {2026}
}

@inproceedings{mu2024large,
  title={Large language models offer an alternative to the traditional approach of topic modelling},
  author={Mu, Yida and Dong, Chun and Bontcheva, Kalina and Song, Xingyi},
  booktitle={Proceedings of the 2024 joint international conference on computational linguistics, language resources and evaluation},
  pages={10160--10171},
  year={2024},
  url          = {https://aclanthology.org/2024.lrec-main.887}
}

@article{mu2024addressing,
  title={Addressing Topic Granularity and Hallucination in Large Language Models for Topic Modelling},
  author={Mu, Yida and Bai, Peizhen and Bontcheva, Kalina and Song, Xingyi},
  journal={arXiv preprint},
  year={2024},
  doi          = {10.48550/ARXIV.2405.00611}
}

@inproceedings{wang2023prompting,
  title={Prompting large language models for topic modeling},
  author={Wang, Han and Prakash, Nirmalendu and Hoang, Nguyen Khoi and Hee, Ming Shan and Naseem, Usman and Lee, Roy Ka-Wei},
  booktitle={2023 IEEE International Conference on Big Data (BigData)},
  pages={1236--1241},
  year={2023},
  organization={IEEE},
  doi          = {10.1109/BIGDATA59044.2023.10386113}
}

@inproceedings{doi2024topic,
  title={Topic modeling for short texts with large language models},
  author={Doi, Tomoki and Isonuma, Masaru and Yanaka, Hitomi},
  booktitle={Proceedings of the 62nd Annual Meeting of the Association for Computational Linguistics (Volume 4: Student Research Workshop)},
  pages={21--33},
  year={2024},
  doi          = {10.18653/V1/2024.ACL-SRW.3}
}

@article{khandelwal2024investigating,
  title={Investigating the Impact of Text Summarization on Topic Modeling},
  author={Khandelwal, Trishia},
  journal={arXiv preprint},
  year={2024},
  doi = {10.48550/arXiv.2410.09063},
}

@inproceedings{rijcken2023towards,
  title={Towards interpreting topic models with ChatGPT},
  author={Rijcken, Emil and Scheepers, Floortje and Zervanou, Kalliopi and Spruit, Marco and Mosteiro, Pablo and Kaymak, Uzay},
  booktitle={The 20th World Congress of the International Fuzzy Systems Association},
  year={2023}
}

@misc{li2023can, 
    title={Can Large Language Models (LLM) label topics from a topic model?}, 
    doi={10.31235/osf.io/23x4m}, 
    publisher={OSF}, 
    author={Li, Dai and Zhang, Bolun and Zhou, Yimang}, 
    year={2023}, 
    month={Jul} 
}

@inproceedings{vaccargiu2024sustainability,
  title={Sustainability in blockchain development: A bert-based analysis of ethereum developer discussions},
  author={Vaccargiu, Matteo and Aufiero, Sabrina and Bartolucci, Silvia and Neykova, Rumyana and Tonelli, Roberto and Destefanis, Giuseppe},
  booktitle={Proceedings of the 28th International Conference on Evaluation and Assessment in Software Engineering},
  pages={381--386},
  year={2024},
  doi          = {10.1145/3661167.3661194}
}

@article{egger2022topic,
  title={A topic modeling comparison between lda, nmf, top2vec, and bertopic to demystify twitter posts},
  author={Egger, Roman and Yu, Joanne},
  journal={Frontiers in sociology},
  volume={7},
  year={2022},
  publisher={Frontiers Media SA},
  DOI={10.3389/fsoc.2022.886498}
}

@inproceedings{sharifian2022analysing,
  title={Analysing longitudinal social science questionnaires: topic modelling with BERT-based embeddings},
  author={Sharifian-Attar, Vida and De, Suparna and Jabbari, Sanaz and Li, Jenny and Moss, Harry and Johnson, Jon},
  booktitle={2022 IEEE international conference on big data},
  pages={5558--5567},
  year={2022},
  organization={IEEE},
  doi          = {10.1109/BIGDATA55660.2022.10020678}
}

@article{abdelrazek2023topic,
  title={Topic modeling algorithms and applications: A survey},
  author={Abdelrazek, Aly and Eid, Yomna and Gawish, Eman and Medhat, Walaa and Hassan, Ahmed},
  journal={Information Systems},
  volume={112},
  pages={102131},
  year={2023},
  publisher={Elsevier},
  doi          = {10.1016/J.IS.2022.102131}
}

@article{grootendorst2022bertopic,
  title={BERTopic: Neural topic modeling with a class-based TF-IDF procedure},
  author={Grootendorst, Maarten},
  journal={arXiv preprint},
  year={2022},
  doi = {10.48550/arXiv.2203.05794},
}

@inproceedings{overney2020not,
  title={How to not get rich: An empirical study of donations in open source},
  author={Overney, Cassandra and Meinicke, Jens and K{\"a}stner, Christian and Vasilescu, Bogdan},
  booktitle={Proceedings of the ACM/IEEE 42nd international conference on software engineering},
  pages={1209--1221},
  year={2020},
  doi          = {10.1145/3377811.3380410}
}

@article{gao2024pyradar,
  title={PyRadar: Towards Automatically Retrieving and Validating Source Code Repository Information for PyPI Packages},
  author={Gao, Kai and Xu, Weiwei and Yang, Wenhao and Zhou, Minghui},
  journal={Proceedings of the ACM on Software Engineering},
  volume={1},
  number={FSE},
  pages={2608--2631},
  year={2024},
  publisher={ACM New York, NY, USA},
  doi          = {10.1145/3660822}
}

@article{zhang2025exploring,
  title={Exploring the Effectiveness of Open-Source Donation Platform: An Empirical Study on Opencollective},
  author={Zhang, Shuoxiao and Tang, Enyi and Gao, Xinyu and Zhang, Zhekai and Shan, Yixiao and Zhang, Haofeng and He, Ziyang and Zhao, Jianhua and Li, Xuandong},
  journal={Journal of Software: Evolution and Process},
  volume={37},
  number={7},
  pages={e70033},
  year={2025},
  publisher={Wiley Online Library},
  doi          = {10.1002/SMR.70033}
}

@article{sauermann2013increasing,
  title={Increasing web survey response rates in innovation research: An experimental study of static and dynamic contact design features},
  author={Sauermann, Henry and Roach, Michael},
  journal={Research Policy},
  volume={42},
  number={1},
  pages={273--286},
  year={2013},
  publisher={Elsevier},
  doi={10.1016/j.respol.2012.05.003}
}

@article{yilmaz2017assessing,
  title={Assessing agreement between raters from the point of coefficients and log-linear models},
  author={Yilmaz, Ayfer Ezgi and Saracbasi, Tulay},
  journal={Journal of Data Science},
  volume={15},
  number={1},
  pages={1--24},
  year={2017},
  doi={10.6339/JDS.201701_15(1).0001}
}

@inproceedings{siska2024examining,
  title={Examining the robustness of LLM evaluation to the distributional assumptions of benchmarks},
  author={Siska, Charlotte and Marazopoulou, Katerina and Ailem, Melissa and Bono, James},
  booktitle={Proceedings of the 62nd Annual Meeting of the Association for Computational Linguistics (Volume 1: Long Papers)},
  pages={10406--10421},
  year={2024},
  doi          = {10.18653/V1/2024.ACL-LONG.560}
}

@article{baltes2022sampling,
  title={Sampling in software engineering research: A critical review and guidelines},
  author={Baltes, Sebastian and Ralph, Paul},
  journal={Empirical Software Engineering},
  volume={27},
  number={4},
  pages={94},
  year={2022},
  publisher={Springer},
  doi          = {10.1007/S10664-021-10072-8}
}

@inproceedings{wagner2024towards,
  title={Towards evaluation guidelines for empirical studies involving llms},
  author={Wagner, Stefan and Bar{\'o}n, Marvin Mu{\~n}oz and Falessi, Davide and Baltes, Sebastian},
  booktitle={2025 IEEE/ACM International Workshop on Methodological Issues with Empirical Studies in Software Engineering (WSESE)},
  pages={24--27},
  year={2025},
  organization={IEEE},
  doi          = {10.1109/WSESE66602.2025.00011}
}

@inproceedings{akbarpour2025unveiling,
  title={Unveiling ruby: Insights from stack overflow and developer survey},
  author={Akbarpour, Nikta and Saleem Mirza, Ahmad and Raoofian, Erfan and Fard, Fatemeh and Rodr{\'\i}guez-P{\'e}rez, Gema},
  booktitle={Proceedings of the 29th International Conference on Evaluation and Assessment in Software Engineering},
  pages={580--591},
  year={2025},
  doi          = {10.1145/3756681.3756955}
}

@article{randolph2005free,
  title={Free-Marginal Multirater Kappa (multirater K [free]): An Alternative to Fleiss' Fixed-Marginal Multirater Kappa.},
  author={Randolph, Justus J},
  journal={Online submission},
  year={2005},
  publisher={ERIC}
}

@article{chaturvedi2015evaluation,
  title={Evaluation of inter-rater agreement and inter-rater reliability for observational data: an overview of concepts and methods},
  author={Chaturvedi, SRBH and Shweta, RC},
  journal={Journal of the Indian Academy of Applied Psychology},
  volume={41},
  number={3},
  pages={20--27},
  year={2015}
}

@misc{log4j2021,
  author       = {{Apache Software Foundation}},
  title        = {{Log4j 2}},
  year         = {2021},
  url          = {https://logging.apache.org/log4j/2.x}
}

@misc{xz2024,
  author       = {{OpenSSF}},
  title        = {{XZ Backdoor (CVE-2024-3094)}},
  year         = {2024},
  url          = {https://openssf.org/blog/2024/03/30/xz-backdoor-cve-2024-3094/}
}

@misc{openssf_scorecard,
  author       = {{OpenSSF}},
  title        = {{OpenSSF Scorecard}},
  year         = {2023},
  url          = {https://github.com/ossf/scorecard}
}

@misc{funding_yml,
  author       = {{GitHub Docs}},
  title        = {{Displaying a sponsor button in your repository}},
  year         = {2023},
  url          = {https://docs.github.com/en/repositories/managing-your-repositorys-settings-and-features/customizing-your-repository/displaying-a-sponsor-button-in-your-repository}
}

@misc{microsoft_foss,
  author       = {{Microsoft}},
  title        = {{FOSS Fund}},
  year         = {2023},
  url          = {https://github.com/microsoft/foss-fund}
}

@misc{stripe_oss,
  author       = {{Stripe}},
  title        = {{Stripe’s commitment to open source on GitHub creates a friendly and empowering experience for all developers.}},
  year         = {2023},
  url          = {https://github.com/customer-stories/stripe}
}

@misc{indeed_foss,
  author       = {{Indeed Engineering}},
  title        = {{The FOSS Contributor Fund: Six Months In}},
  year         = {2019},
  url          = {https://engineering.indeedblog.com/blog/2019/07/foss-fund-six-months-in/}
}

@misc{pypi_metadata_tutorial,
  author       = {{Python Packaging Authority (PyPA)}},
  title        = {{Configuring Metadata}},
  year         = {2025},
  url          = {https://packaging.python.org/en/latest/tutorials/packaging-projects/\#configuring-metadata}
}

@misc{pypa_website,
  author       = {{Python Packaging Authority (PyPA)}},
  title        = {{PyPA Website}},
  year         = {2025},
  url          = {https://www.pypa.io/}
}

@misc{devtools_pypi,
  title        = {devtools},
  author       = {Colvin, Samuel},
  year         = {2023},
  url          = {https://pypi.org/project/devtools/}
}

@misc{pypi_email_issue,
  author       = {{Python Packaging Authority (PyPA)}},
  title        = {{feat: add email for 2FA not yet enabled on upload}},
  year         = {2023},
  url          = {https://github.com/pypi/warehouse/pull/14444}
}

@misc{reimers_sentence_transformers,
  author       = {Nils Reimers and Iryna Gurevych},
  title        = {{Sentence-Transformers: Pretrained Models}},
  year         = {2023},
  url          = {https://www.sbert.net/docs/sentence_transformer/pretrained_models.html}
}

@misc{x_counting_characters,
  author       = {{X Documentation}},
  title        = {{Counting characters when composing Tweets}},
  year         = {2025},
  url          = {https://docs.x.com/fundamentals/counting-characters}
}

@article{verdecchia2025notes,
  title={Notes On Writing Effective Empirical Software Engineering Papers: An Opinionated Primer},
  author={Verdecchia, Roberto and Bogner, Justus},
  journal={ACM SIGSOFT Software Engineering Notes},
  volume={50},
  number={3},
  pages={24--36},
  year={2025},
  publisher={ACM New York, NY, USA},
  doi          = {10.48550/ARXIV.2506.11002}
}

@article{cox2025fifty,
  title={Fifty Years of Open Source Software Supply Chain Security: For decades, software reuse was only a lofty goal. Now it's very real.},
  author={Cox, Russ},
  journal={Queue},
  volume={23},
  number={1},
  pages={84--107},
  year={2025},
  publisher={ACM New York, NY, USA},
  doi          = {10.1145/3722542}
}

@inproceedings{shimada2022githubsponsors,
  title={Github sponsors: exploring a new way to contribute to open source},
  author={Shimada, Naomichi and Xiao, Tao and Hata, Hideaki and Treude, Christoph and Matsumoto, Kenichi},
  booktitle={Proceedings of the 44th international conference on software engineering},
  pages={1058--1069},
  year={2022},
  doi          = {10.1145/3510003.3510116}
}

@inproceedings{zhang2022sponsor,
  title={Who, what, why and how? towards the monetary incentive in crowd collaboration: A case study of github’s sponsor mechanism},
  author={Zhang, Xunhui and Wang, Tao and Yu, Yue and Zeng, Qiubing and Li, Zhixing and Wang, Huaimin},
  booktitle={Proceedings of the 2022 CHI Conference on Human Factors in Computing Systems},
  pages={1--18},
  year={2022},
  doi          = {10.1145/3491102.3501822}
}

@inproceedings{fan2024mygithubsponsors,
  title={"My GitHub Sponsors profile is live!" Investigating the Impact of Twitter/X Mentions on GitHub Sponsors},
  author={Fan, Youmei and Xiao, Tao and Hata, Hideaki and Treude, Christoph and Matsumoto, Kenichi},
  booktitle={Proceedings of the IEEE/ACM 46th International Conference on Software Engineering},
  pages={1--12},
  year={2024},
  doi          = {10.1145/3597503.3639127}
}

@inproceedings{wu2024large,
  title={A large-scale empirical study of open source license usage: Practices and challenges},
  author={Wu, Jiaqi and Bao, Lingfeng and Yang, Xiaohu and Xia, Xin and Hu, Xing},
  booktitle={Proceedings of the 21st International Conference on Mining Software Repositories},
  pages={595--606},
  year={2024},
  doi          = {10.1145/3643991.3644900}
}

@inproceedings{vu2021py2src,
  title={Py2src: Towards the automatic (and reliable) identification of sources for pypi package},
  author={Vu, Duc-Ly},
  booktitle={2021 36th IEEE/ACM International Conference on Automated Software Engineering (ASE)},
  pages={1394--1396},
  year={2021},
  organization={IEEE},
  doi          = {10.1109/ASE51524.2021.9678526}
}

@inproceedings{vu2021lastpymile,
  title={Lastpymile: identifying the discrepancy between sources and packages},
  author={Vu, Duc-Ly and Massacci, Fabio and Pashchenko, Ivan and Plate, Henrik and Sabetta, Antonino},
  booktitle={Proceedings of the 29th ACM Joint Meeting on European Software Engineering Conference and Symposium on the Foundations of Software Engineering},
  pages={780--792},
  year={2021},
  doi          = {10.1145/3468264.3468592}
}

@inproceedings{goswami2020investigating,
  title={Investigating the reproducibility of npm packages},
  author={Goswami, Pronnoy and Gupta, Saksham and Li, Zhiyuan and Meng, Na and Yao, Daphne},
  booktitle={2020 IEEE International Conference on Software Maintenance and Evolution (ICSME)},
  pages={677--681},
  year={2020},
  organization={IEEE},
  doi          = {10.1109/ICSME46990.2020.00071}
}

@article{feng2026charting,
  title={Charting Uncertain Waters: A Socio-Technical Roadmap for Sustaining Open Source Communities in the Age of GenAI},
  author={Feng, Zixuan and Milewicz, Reed and Murphy-Hill, Emerson and Menezes, Tyler and Serebrenik, Alexander and Steinmacher, Igor and Sarma, Anita},
  journal={ACM Transactions on Software Engineering and Methodology},
  year={2026},
  publisher={ACM New York, NY},
  doi = {10.1145/3789210}
}

@misc{litellm,
  title        = {PYSEC-2026-2: Malicious code in litellm (PyPI)},
  author       = {{Open Source Vulnerabilities (OSV)}},
  year         = {2026},
  url          = {https://osv.dev/vulnerability/PYSEC-2026-2}
}
